\documentstyle[11pt]{article}
\begin{document} 
\input{epsf}


\title{\LARGE \bf Cosmic Strings in the Abelian Higgs Model with Conformal
Coupling to Gravity}


\author{ Y. Verbin$^\star $\\
Department of Natural Sciences, \\ The Open University of Israel, 
P.O.B. 39328, Tel Aviv 61392, Israel\\}

\date{ }

\bibliographystyle{unsrt}
\maketitle


\begin{abstract} Cosmic string solutions of the abelian Higgs model with
conformal coupling to gravity are shown to exist. The main characteristics of
the solutions are presented and the differences with respect to the minimally
coupled case are studied. 
An important difference is the absence of Bogomolnyi cosmic string solutions 
for conformal coupling.
Several new features of the abelian Higgs cosmic
strings of both types are discussed. The most interesting is perhaps a relation
between the angular deficit and the central magnetic field which is
bounded by a critical value.\\
{\em PACS numbers: 11.27.+d, 04.20.Jb, 04.40.Nr}
\end{abstract}

\vskip2pc

\renewcommand{\theequation}{\arabic{section}.\arabic{equation}}
\setcounter{equation}{0}
\section{Introduction}

Cosmic strings \cite{vilsh,KibbleH} were introduced into cosmology by 
Kibble  \cite{Kibble1}, Zel'dovich \cite{Zel} and Vilenkin \cite{Vil1} as
(linear) topological  defects which may have been formed during phase
transitions in the early  universe. Cosmic strings are considered as possible
sources for density perturbations and hence for structure formation in the
universe.  

The simplest geometry of a cosmic string is that of a straight infinite one, 
which may represent anyhow a generic cosmic string from close up. The straight
infinite (and static) string is therefore a source of a static and cylindrically
symmetric gravitational field.
  
It is well-known \cite{exact-sol} that the general
static cylindrically-symmetric vacuum solution of Einstein equations is
characterized by two free parameters. A simple representation of this is the
so-called Kasner solution:  

\begin{eqnarray}
ds^{2} = (kr)^{2a}dt^{2} - (kr)^{2c}dz^{2} - dr^{2}
- \beta^{2} (kr)^{2(b-1)}r^{2}d{\phi}^2, \label{Kasner1}
\end{eqnarray}
where $k$ sets the length scale, $\beta$ will be discussed below and $a$, $b$, 
$c$ satisfy the Kasner conditions:
\begin{equation}
a + b + c=a^2 + b^2 + c^2=1. \label{Kasner2}
\end{equation}

A question of fundamental interest is the interpretation of these parameters
and the connection between them and the internal properties of the matter
distribution, i.e. of the cosmic string in the present context.

\vspace{2 mm}
\line(1,0){210}
\vspace{4 mm}

\noindent $^\star$Electronic address: verbin@oumail.openu.ac.il
\newpage

The simplest model for a cosmic string is the so called "gauge string"
\cite{KibbleH} which is an idealized cylindrical mass distribution with a
finite radial extension having ${\cal T}^0_0={\cal T}^3_3$ as the only
non-vanishing components of the energy-momentum tensor. It is well-known by now
that this string picks up from the Kasner family a very simple solution namely
$a=c=0, b=1$ which is evidently locally flat. However, there is a global
non-trivial effect i.e. the geometry is that of a cone. The parameter $\beta$
represents a conic angular deficit \cite{Marder2,Bonnor} which is also related
to the mass distribution of the source.

The explicit relation between the
angular deficit $\delta\phi = 2 \pi (1-\beta)$ and the "inertial mass" (per
unit length) $\tilde{m}$, of a gauge string was found to be
\begin{equation}
\delta\phi = 8 \pi G\tilde{m},
\label{angdef}
\end{equation}
first \cite{Vil1} in the linearized approximation assuming also an
infinitesimally thin source, and then \cite{Gott,Hiscock} by solving the full
non-linear Einstein equations around a uniform source (constant ${\cal T}^0_0$)
with a finite radius. The same relation was also derived \cite{Linet1} for a
non-uniform source. Since the space-time around a gauge string is locally flat,
this angular deficit is the only geometrical evidence of its existence.

Further study of the subject \cite{Garfinkle1} involved a more realistic model
(i.e. the abelian Higgs model) for the cosmic strings and the analysis of the
full coupled field equations for the gravitational field and matter (scalar +
vector) fields. It became clear that (\ref{angdef}) is only an approximation
and several corrections were calculated
\cite{LagMatz,GarfinkleLag,LagGarfinkle,CNV}. 

The cosmic strings of the abelian Higgs model are the simplest and most studied
ones.  If however, gravity is a scalar-tensor theory as suggested by string
theory, one should examine the existence of cosmic strings in this case as
well. This has been done already for Brans-Dicke theory and its extensions
\cite{GO, PM, Guimaraes} and for dilatonic gravity \cite{GregorySantos} where
an extra scalar field (dilaton) couples to the Higgs system and to the metric
tensor. Cosmic string solutions were found and their characteristics were
discussed.

There exists yet another simple system which may give rise to cosmic strings
where gravity is purely tensorial but couples non-minimally to the matter
fields. It is the abelian Higgs model where the charged scalar field has a
(non-minimal) "conformal" coupling to gravity. It seems that cosmic strings of
this model have never been discussed in the literature, although it is not much
more complicated than the minimally coupled abelian Higgs model. This paper
contains therefore a verifications (by construction) of the existence of such
solutions and a study of their characteristics. In order to have a basis for
appreciating the new results, the next section (2) concentrates in cosmic
strings in the minimally coupled case. The most important features are
presented - most of them well-known but some interesting novelties are added.
Section 3 contains the general analysis of the conformally coupled case and
section 4 contains numerical results which establish the  existence of the new
kind of cosmic strings and a discussion of their main features.  Generally,
they are found to be similar to the minimally coupled ones.

\section{Abelian Higgs Cosmic Strings}
\setcounter{equation}{0}
The action which allows self-gravitating abelian flux tubes is

\begin{equation}
        S = \int d^4 x \sqrt{\mid g\mid } \left({1\over 2}D_{\mu}\Phi ^
        {\ast}D^{\mu}\Phi 
        - {{\lambda  }\over 4}(\Phi ^{\ast} \Phi - v^2)^2 -
        {1\over 4}{F}_{\mu \nu}{F}^{\mu \nu} + \frac{1}{16\pi G} 
        {\cal R}\right)
\label{higgsaction}
\end{equation}
where ${\cal R}$ is the Ricci scalar, ${F}_{\mu \nu}$ the abelian field
strength, $\Phi$ is a complex scalar field and $D_\mu = \nabla _{\mu} -
ieA_{\mu}$ is the usual gauge covariant  derivative. We use units in which
$\hbar=c=1$.

Because of the cylindrical symmetry of the source we will use a line element 
of the form:
\begin{equation}
ds^{2} = N^{2}(r)dt^{2} - d{r}^{2} - L^{2}(r)d{\phi}^2 - K^{2}(r)dz^{2}
\label{lineelement}
\end{equation}
and the usual Nielsen-Olesen ansatz for the +1 flux unit:
\begin{eqnarray}
\Phi=vf(r)e^{i\phi}, \hspace{10 mm}
A_\mu dx^\mu = {1\over e}(1-P(r))d\phi .
\label{NOansatz}
\end{eqnarray}
This gives rise to the following field equations for the Abelian Higgs flux 
tube:
\begin{eqnarray}
\frac{(NKLf')'}{NKL} + \left(\lambda v^2 (1-f^2) - 
\frac{P^2}{L^2}\right)f = 0 \\
\frac{L}{NK}\left(\frac{NK}{L} P'\right)' - e^2 v^2 f^2 P = 0 \label{fluxtube}
\end{eqnarray}
With the line element (\ref{lineelement}) the components of the Ricci tensor 
are:
\begin{eqnarray}
{\cal R}^{0}_{0} = - \frac{(LKN')'}{NLK}  ,\hspace{8 mm}
{\cal R}^{r}_{r} = - \frac{N''}{N} - \frac{L''}{L} - \frac{K''}{K}  
,\hspace{8 mm}
{\cal R}^{\phi}_{\phi} = - \frac{(NKL')'}{NLK} ,\hspace{8 mm}
{\cal R}^{z}_{z} = - \frac{(NLK')'}{NLK}  .\label{Ricci}
\end{eqnarray}

The source is described by the energy-momentum tensor with the
following components:
\begin{eqnarray}
     {\cal T}^{0}_{0} &=& \rho = \varepsilon _s +\varepsilon _v +
     \varepsilon _{sv} + u  \nonumber \\
     {\cal T}^{r}_{r} &=& -p_r = -\varepsilon _s -\varepsilon _v +
     \varepsilon _{sv} + u \nonumber \\
     {\cal T}^{\phi}_{\phi} &=& -p_{\phi} = \varepsilon _s -
     \varepsilon _v -\varepsilon_{sv} + u \nonumber \\
     {\cal T}^{z}_{z} &=& -p_z = \rho 
 \label{Tmunu}
 \end{eqnarray}
where
\begin{eqnarray}
\varepsilon_s = {v^2 \over 2} f'^2 ,\hspace{10 mm}
\varepsilon_v = \frac{P'^2}{2e^2 L^2} ,\hspace{10 mm} 
\varepsilon_{sv} = \frac{v^2 P^2 f^2}{2L^2} ,\hspace{10 mm}
u = \frac{\lambda v^4}{4} (1-f^2)^2.
\label{densities}
\end{eqnarray}
For further use we define the magnetic field as:
\begin{equation}
{\cal B}  =  -\frac{P'}{eL}.
\label{magn}
\end{equation}

It turns out to be convenient to use Einstein equations in the form:
\begin{equation}
{\cal R}_{\mu\nu} = -8 \pi G({\cal T}_{\mu \nu} -
\frac{1}{2} g_{\mu \nu}{\cal T})
\label{EinsteinR}
\end{equation}
where ${\cal T}$ is the contracted energy-momentum tensor:
${\cal T} = \rho - p_z - p_r - p_{\phi }.$

By insertion of ${\cal T}_{\mu\nu}$ one obtains:
\begin{eqnarray}
\frac{(LKN')'}{NLK} &=& 4 \pi G(\rho +p_r+p_{\phi} +p_z)=
8 \pi G(\varepsilon_v - u) 
\label{Einst0} 
\end{eqnarray}
\begin{eqnarray}
\frac{(NKL')'}{NLK} &=& -4 \pi G(\rho -p_r+p_{\phi} -p_z)
= - 8 \pi G(\varepsilon_v + 2 \varepsilon_{sv} + u) 
\label{Einstphi}
\end{eqnarray}
\begin{eqnarray}
\frac{(LNK')'}{NLK} &=& -4 \pi G(\rho -p_r-p_{\phi} +p_z)=
8 \pi G(\varepsilon_v - u)
\label{Einst3}
\end{eqnarray}
and instead of the "radial" part of (\ref{EinsteinR}) we take the following 
combination:
\begin{eqnarray}
\frac{N'}{N} \frac{L'}{L} + \frac{L'}{L} \frac{K'}{K}+
\frac{K'}{K} \frac{N'}{N} = 8 \pi G p_r = 
8 \pi G(\varepsilon _s +\varepsilon _v - \varepsilon _{sv} - u),
\label{constraint}
\end{eqnarray}
which is not an independent equation but serves as a constraint. In vacuum the
right-hand-sides of these equations vanish and the first three of them are
trivially integrated. In this way we may get back Kasner's line element 
(\ref{Kasner1}). This is therefore the asymptotic form of the metric tensor
around any (transversally) localized source and especially around an abelian
Higgs flux tube.

Moreover, it is easy to get convinced that due to the symmetry under boosts
along the string axis, $K=N$.

The equations become more transparent if we express all lengths in terms of the
scalar characteristic length scale ($1/\sqrt{\lambda v^2}$, the "correlation 
length" in the superconductivity terminology). We therefore change to the 
dimensionless length
coordinate $x=\sqrt{\lambda v^2}r$ and metric component $\sqrt{\lambda v^2}L$
which we still denote by $L$. We also introduce the two parameters 
$\alpha=e^2/\lambda$ and $\gamma=8\pi Gv^2$. In terms of these new quantities we
get a two parameter system of 4 coupled non-linear ordinary differential
equations (Now of course $'=d/dx$):

\begin{eqnarray}
\frac{(N^2 Lf')'}{N^2 L} + \left(1-f^2 - \frac{P^2}{L^2}\right)f = 0   
\label{systemNO1}\\
\frac{L}{N^2}\left(\frac{N^2 P'}{L}\right)' - \alpha f^2 P = 0     
\label{systemNO2}
\end{eqnarray}
\begin{eqnarray}
\frac{(LNN')'}{N^2 L} &=&\gamma\left(\frac{P'^2}{2\alpha L^2} - \frac{1}{4}
(1-f^2)^2\right)  
\label{systemE1}\\
\frac{(N^2 L')'}{N^2 L}
&=& - \gamma\left(\frac{P'^2}{2\alpha L^2} + \frac{P^2 f^2}{L^2} + \frac{1}{4}
(1-f^2)^2\right).
\label{systemE2}
\end{eqnarray}
We have also to keep in mind the existence of the constraint (\ref{constraint}) 
which gets the
following form:
\begin{eqnarray}
\frac{N'}{N} \left(2\frac{L'}{L} + \frac{N'}{N}\right) = 
\gamma\left(\frac{f'^2}{2} + \frac{P'^2}{2\alpha L^2} -
\frac{P^2 f^2}{2L^2} - \frac{1}{4}(1-f^2)^2 \right) .
\label{constraintmod}
\end{eqnarray}

In order to get string solutions the scalar and gauge field should satisfy the 
following boundary conditions:
\begin{eqnarray}
f(0)=0, \hspace{2cm} \lim_{x\rightarrow \infty} f(x) = 1 \nonumber \\
P(0)=1, \hspace{2cm} \lim_{x\rightarrow \infty} P(x) = 0
\label{boundarycond}
\end{eqnarray}
and regularity of the geometry on the symmetry axis will be guarantied by the
"initial conditions":
\begin{eqnarray}
L(0)=0, \hspace{2cm} L'(0) = 1 \nonumber \\
N(0) = 1, \hspace{2cm} N'(0) = 0
\label{initcond}
\end{eqnarray}

It is well-known that even in flat space these field equations can only be
solved numerically. However, much can be said about the nature of the solutions
even without explicitly solving the field equations
\cite{Garfinkle1,LagGarfinkle,FIU,CNV}. The first is that the abelian Higgs
string has a vanishing gravitational mass which means that the spacetime around
the string is locally flat except in its core. However, there is a non-trivial
global effect namely a conical structure of the space which is quantified by an
angular deficit. This does not saturate all the possibilities. There are
further types of solutions of (\ref{systemNO1})-(\ref{systemE2}) with the same
boundary (and "initial") conditions which are not asymptotically  flat but have
interesting physical interpretation. Thus, a point in the $\alpha$-$\gamma$
plane does not always represent a single solution. The various solutions are
distinguished by their asymptotic geometries. Here we will concentrate in the
cosmic string solutions, i.e. those which are asymptotically conical.

For analyzing the solutions and obtaining the above-mentioned features and some
additional ones, we 
introduce the Tolman mass (per unit length) of this system, $M$:
\begin{eqnarray}
GM&=&2\pi G\int _{0}^\infty dr N^2 L(\rho +p_r+p_{\phi} +p_z)=\nonumber \\
&=&\frac{\gamma}{2} \int _{0}^\infty dx 
N^2 L \left(\frac{P'^2}{2\alpha L^2} - \frac{1}{4} (1-f^2)^2\right) = 
\frac{1}{2} \lim_{x\rightarrow\infty} (LNN')
\label{mass}
\end{eqnarray}
where the last equality comes from the "temporal" Einstein equation. 

The string angular deficit will be related to the "angular" Einstein equation by
defining:
\begin{eqnarray}
GW&=&-2\pi G\int _{0}^\infty dr N^2 L(\rho -p_r+p_{\phi} -p_z)=\nonumber \\
&=&-\frac{\gamma}{2} \int _{0}^\infty dx 
N^2 L \left(\frac{P'^2}{2\alpha L^2} + \frac{P^2 f^2}{L^2} + \frac{1}{4} 
(1-f^2)^2\right) = 
\frac{1}{2} (\lim_{x\rightarrow\infty} (N^2L') - 1)
\label{GW}
\end{eqnarray}
It is assumed of course that the integrals converge and the limits exist. This
will be verified numerically later.

The asymptotic form of the metric tensor is easily found by direct integration
of the two Einstein equations (\ref{systemE1})-(\ref{systemE2}) using the
boundary conditions and the definitions of $M$ and $W$. It is  of the Kasner
form  
\begin{eqnarray}
N(x)=K(x)\sim \kappa x^a ,\hspace{10 mm} L(x)\sim \beta x^b , 
\label{asymptKasner1}
\end{eqnarray}
with:
\begin{eqnarray}
a=\frac{2GM}{2G (2M+W)+1} ,\hspace{6 mm}
b=\frac{2GW+1}{2G (2M+W)+1} ,\hspace{6 mm}
\kappa ^2\beta=2G (2M+W)+1.
\label{asymptKasner2}
\end{eqnarray}
The constant $\kappa$ appears free in the asymptotic solution, but it is 
uniquely fixed in the complete one by the boundary conditions on the metric.

Now we notice that the "radial" Einstein equation, (\ref{constraintmod})
can be used at $x\rightarrow\infty$ to give in this case the following relation 
which is equivalent to the quadratic Kasner condition in (\ref{Kasner2}):
\begin{equation}
GM(GM+2GW+1)=0 .
\label{MW}
\end{equation}
This immediately indicates the existence of two branches of solutions. $M=0$
corresponds to   the cosmic string branch since $a=c=0, b=1$. The other turns
out to have non-flat asymptotic  behavior ($a=c=2/3, b=-1/3$), much like that
of the Melvin solution and we may call it the  "Melvin branch". We will not
focus on this branch since it is clear by its asymptotics that it cannot
represent cosmic  strings. We will however refer to it on passing in order to
complete the picture.

In the cosmic string branch one can easily identify the angular deficit 
$\delta\phi$ by:
\begin{equation}
1-\frac{\delta\phi}{2\pi}=L'(\infty) =\frac{2GW+1}
{N^2(\infty)} .
\label{angdefLW}
\end{equation}

It is obvious that $W$ contains contribution not only from the inertial mass
(integral over $\rho$), but also from the pressure terms so eq. (\ref{angdef}),
$\delta\phi = 8 \pi G\tilde{m}$, is valid only in circumstances 
when the transverse pressures are negligible. Otherwise, one should replace
$\tilde{m}$ by $\tilde{m}-\tilde{p}_r$, where $\tilde{p}_r$ is the appropriate
integral over $p_r$ (with $N(\infty)$ absorbed in the measure). In short:
\begin{equation}
\tilde{m}=2\pi \int _{0}^\infty dr N^2 L\rho/N^2(\infty) , \hspace{6 mm}
\tilde{p}_r=2\pi \int _{0}^\infty dr N^2 Lp_r/N^2(\infty)
\label{imasspr}
\end{equation}
Recall also that $M=0$. Moreover, it is inconsistent to take
$\kappa=N(\infty)=1$ since we have already made the choice $N(0)=1$ in order to
be able to treat on the same footing the solutions in both branches. This
factor represents a red/blue-shift of time between infinity and the core of the
string and adds its own contribution to the angular deficit. We have therefore
a generalized Vilenkin relation:
\begin{equation}
\frac{\delta\phi}{2\pi}=4G(\tilde{m}-\tilde{p}_r)+1-\frac{1}{N^2(\infty)} .
\label{angdefVil2}
\end{equation} 

Actually this result does not depend upon the detailed properties of the model
and is valid for any static  matter distribution with cylindrical symmetry and
boost symmetry along the string axis, provided it is localized enough so the
relevant integrals converge. In the case of the minimally coupled abelian Higgs
model one may get a more explicit expression by use of eqs.
(\ref{Tmunu})-(\ref{densities}).      

The angular deficit may be written in an alternative form if we start from the
following identity:
\begin{eqnarray}
\frac{L''}{L}=(\frac{N'}{N})^2 - 4 \pi G(2\rho +p_r+p_{\phi})
\label{Einstphi2}
\end{eqnarray}
which is obtained for $\rho +p_z=0 \hspace{3 mm} (\Rightarrow K=N)$ from 
(\ref{Einstphi}) and (\ref{constraint}). If we now multiply both sides by 
$L(r)$ and integrate we find after some manipulations (integrations by parts 
etc.) and use of $M=0$ that:
\begin{equation}
\frac{\delta\phi}{2\pi}=8\pi G\int _{0}^\infty dr L(\rho +
{1\over 4}(p_r+p_{\phi})).
\label{angledef3}
\end{equation}
Note that the measure here is a purely two-dimensional one which is related to
the induced metric of the constant time and constant $z$ surface. A
"predecessor" of this relation was derived already by Garfinkle and Laguna
\cite{Garfinkle1,LagGarfinkle} although in a form where only the $\rho$ term of
(\ref{angledef3}) was explicit and the second term was expressed in terms of
the metric components. This result is also model-independent in the same sense
as explained above.  For the abelian Higgs model we obtain the following
explicit expression:
\begin{equation}
\frac{\delta\phi}{2\pi}=4\pi G\int _{0}^\infty dr L(2\varepsilon _s +
3\varepsilon _v + 2\varepsilon _{sv} +u).
\label{angledef3aH}
\end{equation}
   
An additional relation unnoticed so far is a simple connection between the
angular deficit of an abelian Higgs string and the central value of the
magnetic field ${\cal B}(0)$. One may prefer to use the dimensionless
magnetic field $B(x)=-P'(x)/L(x)=(8\pi G /e)  \alpha {\cal B} / \gamma$.
At any rate the relation is: 
\begin{equation}
\frac{\delta\phi}{2\pi} =1-\frac{1-\frac{\gamma}{\alpha} B(0)}{N^2(\infty)} 
=1-\frac{1-\frac{8\pi G}{e}{\cal B}(0)}{N^2(\infty)} .
\label{angdefB}
\end{equation}
This result is obtained from the following identity which amounts to an
integration of one combination (i.e. the  difference) of the two equations
(\ref{systemE1})-(\ref{systemE2}):
\begin{equation}
\frac{d}{dx}\left[N^2 L\left(\frac{N'}{N}- \frac{L'}{L} - \frac{\gamma}{\alpha}
\frac{PP'}{L^2}\right) \right]=0 .
\label{firstordercomb}
\end{equation}
If we use this identity for the string branch at both $x\rightarrow 0$ and
$x\rightarrow\infty$ we get the following which directly leads from 
(\ref{angdefLW}) to (\ref{angdefB}):
\begin{equation}
G W=-\frac{\gamma}{2 \alpha} B(0).
\label{WB2}
\end{equation}
This relation is easily generalized for the Melvin branch:
\begin{equation}
G (M-W)=\frac{\gamma}{2 \alpha} B(0).
\label{WBMelvin}
\end{equation}
Since now $M\neq 0$ we may use the relation (\ref{MW}) to solve for the mass 
and the $W$-parameter in terms of the single combination $\gamma B(0)/\alpha$:
\begin{eqnarray}
G M= \frac{\gamma}{3\alpha} B(0)-\frac{1}{3}\\
G W=-\frac{\gamma}{6\alpha} B(0)-\frac{1}{3}.
\label{MWdeponB}
\end{eqnarray}
Note that in the case of the -1 flux unit the magnetic field changes sign so 
$B(0)$ should actually be replaced in the above equations by $|B(0)|$ in order 
to cover both cases.

As numerical analysis shows (sec. 4), the angular deficit increases with
$\gamma$ and decreases with $\alpha$. An analytic demonstration for the
$\gamma$ dependence is obtained in the  simplest solution for this system,
namely the Bogomolnyi solution \cite{Linet2,GibbCom}  which  exists for
$\alpha=2$. 
In this case the system is in a state where the repulsion due to vector exchange
is exactly balanced by the attraction due to the scalar exchange.
This is reflected by an identically vanishing gravitational potential, namely,
$N(x)=1$. The
field equations turn out to be equivalent to  the following first order
equations:
\begin{eqnarray}
Lf'-Pf&=&0  \\
P'+L(1-f^2)&=&0  \label{bog2} \\ 
L'-\frac{\gamma}{2} P(1-f^2)&=&1-\frac{\gamma}{2}.
\label{bog3}
\end{eqnarray}
Using e.g. (\ref{angdefLW}) and (\ref{WB2}) with $B(0)=1$ (which follows from
(\ref{bog2})),  we find a simple expression for the angular deficit which was
already noticed by Linet \cite{Linet2}:
\begin{equation}
\delta \phi = \pi \gamma .
\label{angdefbog}
\end{equation}
This linear increase is evidently associated with a maximal value of $\gamma =
2$. Beyond this value the solutions are closed, namely, outside the core of the
string, $L(x)$ decreases linearly and vanishes at a finite distance from the 
string axis. The geometry is still conic but of an inverted one where the apex
of the cone  is at the point where $L=0$ \cite{Gott,Ortiz}. 

If we now return to arbitrary values of $\alpha$ and $\gamma$ and look back at 
(\ref{angdefB}) we find a critical value for
the central magnetic field of an abelian Higgs cosmic string: 
\begin{equation}
{\cal B}_{c}=e/8\pi G  ,\hspace{10 mm} B_{c}=\alpha/\gamma. 
\label{critB}
\end{equation} 
Note that for a given $e$, say the electromagnetic value, the value of the 
critical magnetic  field, ${\cal B}_{c}$ is unique. However, the dimensionless
critical magnetic field is given in terms of both $\alpha$ and $\gamma$ so
there is a whole curve in the $\alpha$-$\gamma$ plane which corresponds to
solutions with the same critical asymptotic behavior.  

\section{Abelian Higgs Cosmic Strings with Non-Minimal Coupling to Gravity}
\setcounter{equation}{0}

The possibility of non-minimal coupling ($R\Phi^2$) between gravity and scalar 
fields was investigated by many authors and the special coefficient of 1/12 was
found to be of special interest due to the conformal symmetry associated with
such a coupling \cite{Penrose,CCJ,Bek,BirrDav}. It is however quite surprising
that the possibility of cosmic strings in this system has never been discussed
in the literature. The action contains thus an additional term for the
non-minimal "conformal coupling":

\begin{equation}
      S = \int d^4 x \sqrt{\mid g\mid } \left({1\over 2}D_{\mu}\Phi ^
      {\ast}D^{\mu}\Phi 
      - {{\lambda  }\over 4}(|\Phi|^2 - v^2)^2 - {1\over 12}{\cal R}|\Phi|^2 -
      {1\over 4}{F}_{\mu \nu}{F}^{\mu \nu} + \frac{1}{16\pi G} 
      {\cal R}\right).
\label{higgsactioncc}
\end{equation}
The modification in the abelian Higgs sector is only a "renormalization" of the 
self interaction coupling constant $\lambda$ to $(1-\gamma /6)\lambda$ which 
results from the identity
\begin{equation}
{\cal R}=8\pi G{\cal T}=8\pi Gv^2 \lambda (v^2-|\Phi|^2)
\label{RTPHI}
\end{equation}
On the other hand there are considerable modifications in the Einstein 
equations due to the new energy-momentum tensor:
\begin{equation}
{\cal T}_{\mu\nu} = {\cal T}_{\mu\nu}^{(minimal)}+{1\over 6}
\left (g_{\mu \nu} \nabla ^{\rho }\nabla
{\rho } |\Phi|^2 - \nabla _{\mu }\nabla_{\nu } |\Phi|^2  - 
{\cal G}_{\mu \nu} |\Phi|^2 \right)
\label{confTmn1}
\end{equation}
or by using Einstein equation:
\begin{equation}
{\cal T}_{\mu\nu} = \frac{1}{1-{\gamma\over 6}|\Phi|^2}\left [ 
{\cal T}_{\mu\nu}^{(minimal)}+{1\over 6}
\left (g_{\mu \nu} \nabla ^{\rho }\nabla
_{\rho } |\Phi|^2 - \nabla _{\mu }\nabla_{\nu } |\Phi|^2 \right)\right].
\label{confTmn2}
\end{equation}
${\cal T}_{\mu\nu}^{(minimal)}$ is of course the energy-momentum tensor of the 
minimally coupled system of the previous section.

The same cosmic string ansatz gives now the following form of
Einstein equations:
\begin{eqnarray}
\frac{(LKN')'}{NLK} = \frac{8 \pi G}{1-{\gamma\over 6} f^2}
\left({1\over 3}\varepsilon _s +
\varepsilon _v + {1\over 3}\varepsilon _{sv} - w + 
\frac{v^2}{3}\frac{N'}{N}ff'  \right)
\label{Einst0c}
\end{eqnarray}
\begin{eqnarray}
\frac{(NKL')'}{NLK} = \frac{8 \pi G}{1-{\gamma\over 6} f^2}
\left({1\over 3}\varepsilon _s -\varepsilon _v - {5\over 3}\varepsilon _{sv} - 
w + \frac{v^2}{3}\frac{L'}{L}ff'  \right)
\label{Einstphic}
\end{eqnarray}
\begin{eqnarray}
\frac{(LNK')'}{NLK} = \frac{8 \pi G}{1-{\gamma\over 6} f^2}
\left({1\over 3}\varepsilon _s +\varepsilon _v + {1\over 3}\varepsilon _{sv} - 
w + \frac{v^2}{3}\frac{K'}{K}ff'  \right)
\label{Einst3c}
\end{eqnarray} 
where we use the field equation for the scalar and define the combination:
\begin{equation}
w = \frac{\lambda v^4}{4} (1-f^2)
\left(1-{1\over 3}(1+{\gamma\over 3})f^2\right).
\end{equation}
The radial constraint is now:
\begin{eqnarray}
\frac{N'}{N} \frac{L'}{L} + \frac{L'}{L} \frac{K'}{K}+ 
\frac{K'}{K} \frac{N'}{N} = 
\frac{8 \pi G}{1-{\gamma\over 6} f^2}\left(\varepsilon _s +\varepsilon _v - 
\varepsilon _{sv} - u +\frac{v^2}{3}\frac{(NKL)'}{NKL}ff' \right).
\label{constraintcc}
\end{eqnarray}

A glance at the Einstein equations for this case reveals that their structure
is quite similar to that for minimal coupling. The main difference is the
appearance of a "field dependent gravitational constant",  
$G/(1-{\gamma \over 6} f^2)$ 
because of which we may expect more pronounced gravitational effects, since it
is larger than $G$. Moreover,  there seems to be a critical value of the
scalar vacuum expectation value $v$ since $\gamma = 6$  is singular as seen for
example from the "renormalization" of $\lambda$ and from other results that
will follow. We will thus concentrate in the region $\gamma < 6$.

By subtracting (\ref{Einst3c}) from (\ref{Einst0c}) one can show that the
boost symmetry along the string axis is still valid hence $K=N$. We also note
that the last term in the right-hand side in each of
(\ref{Einst0c})-(\ref{Einst3c}) may be transferred to the left-hand-side to
form a total derivative so we write:

\begin{eqnarray}
\left((1-{\gamma\over 6} f^2)LNN'\right)' = 8 \pi G N^2 L\left({1\over 3}
\varepsilon _s + \varepsilon _v + {1\over 3}\varepsilon _{sv} - w \right)
\label{Einst0ci}
\end{eqnarray}
\begin{eqnarray}
\left((1-{\gamma\over 6} f^2)N^2 L'\right)' = 8 \pi G N^2 L\left({1\over 3}
\varepsilon _s - \varepsilon _v - {5\over 3}\varepsilon _{sv} - w \right).
\label{Einstphici}
\end{eqnarray}
It is also useful to write the equations explicitly in terms of the 
dimensionless variables:
\begin{eqnarray}
\left((1-{\gamma\over 6} f^2)LNN'\right)' = \gamma N^2 L\left({f'^2\over 6} +
     \frac{P'^2}{2\alpha L^2} + \frac{P^2 f^2}{6L^2} - 
     {1\over 4}(1-f^2)\left(1-{1\over 3}
     (1+{\gamma\over 3})f^2\right) \right)
\label{Einst0cidim}
\end{eqnarray}
\begin{eqnarray}
\left((1-{\gamma\over 6} f^2)N^2 L'\right)' = \gamma N^2 L\left({f'^2\over 6} 
-\frac{P'^2}{2\alpha L^2} - {5\over 6}\frac{P^2 f^2}{L^2} - 
{1\over 4}(1-f^2)\left(1-{1\over 3}(1+{\gamma\over 3})f^2\right) \right).
\label{Einstphicidim}
\end{eqnarray}

A first order equation analogous to (\ref{firstordercomb}) is obtained
from 
the difference (\ref{Einst0cidim})-(\ref{Einstphicidim}):
\begin{equation}
\frac{d}{dx}\left[N^2 L\left((1-{\gamma\over 6} f^2)\left(\frac{N'}{N}- 
\frac{L'}{L}\right) - 
\frac{\gamma}{\alpha} \frac{PP'}{L^2}\right)\right] =0 .
\label{firstordercombcc}
\end{equation}
The constraint (\ref{constraintcc}) gets the form
\begin{eqnarray}
\frac{N'}{N} \left(2\frac{L'}{L} + \frac{N'}{N}\right) = 
\frac{\gamma}{1-{\gamma\over 6} f^2}\left(\frac{f'^2}{2} + 
\frac{P'^2}{2\alpha L^2} -\frac{P^2 f^2}{2L^2} - 
\frac{1}{4}(1-f^2)^2+\frac{1}{3}\frac{(N^2 L)'}{N^2 L}ff' \right).
\label{constraintccdim}
\end{eqnarray}

Integration of (\ref{Einst0cidim}) and (\ref{Einstphicidim}) gives the 
following integral expressions for $M$ and $W$:
\begin{eqnarray}
GM&=& \frac{1}{2} \lim_{x\rightarrow\infty} (LNN')=\nonumber \\
&=&\frac{\gamma}{2(1-{\gamma\over 6})} \int _{0}^\infty dx 
N^2 L \left({f'^2\over 6} +
     \frac{P'^2}{2\alpha L^2} + \frac{P^2 f^2}{6L^2} - 
     {1\over 4}(1-f^2)\left(1-{1\over 3}
     (1+{\gamma\over 3})f^2\right) \right)
\label{GMcc}
\end{eqnarray}
\begin{eqnarray}
GW&=&\frac{1}{2} (\lim_{x\rightarrow\infty} (N^2L') - 1)=\nonumber \\
&=&\frac{\gamma}{2(1-{\gamma\over 6})} \int _{0}^\infty dx 
N^2 L \left({f'^2\over 6} -
     \frac{P'^2}{2\alpha L^2} - {5\over 6}\frac{P^2 f^2}{L^2} - 
     {1\over 4}(1-f^2)\left(1-{1\over 3}
     (1+{\gamma\over 3})f^2\right) \right).
\label{GWcc}
\end{eqnarray}
The asymptotic form of the metric tensor is still described by
(\ref{asymptKasner1})-(\ref{MW}) and the angular deficit is still given by
(\ref{angdefLW}) or (\ref{angdefVil2}) provided we use the appropriate
energy-momentum tensor. Since (\ref{angledef3}) is also still valid, an 
expression analogous to (\ref{angledef3aH}) may be immediately  found:
\begin{equation}
\frac{\delta\phi}{2\pi}=4\pi G\int _{0}^\infty dr 
\frac{L}{1-{\gamma\over 6} f^2}
\left( \varepsilon _s +3\varepsilon _v + \varepsilon _{sv} +
\frac{\lambda v^4}{4} (1-f^2)\left(1+(1-{\gamma\over 3})f^2\right) 
+v^2\frac{N'}{N}ff' \right)
\label{angledef3aHcc}
\end{equation}
   
A relation between the angular deficit and the magnetic field exists here 
as well. We can use eq. (\ref{firstordercombcc}) to obtain:
\begin{equation}
(1-{\gamma\over 6})G (M-W)=\frac{\gamma}{2 \alpha} B(0)-{\gamma\over 12}.
\label{WBcc}
\end{equation}
From here we find that for the cosmic string branch 
\begin{equation}
GW=-\frac{\frac{\gamma}{2 \alpha} B(0)-{\gamma\over 12}}{1-{\gamma\over 6}}
\label{WB2cc}
\end{equation}
and
\begin{equation}
\frac{\delta\phi}{2\pi} =
1-\frac{1-\frac{\gamma}{\alpha} B(0)}{(1-{\gamma\over 6})N^2(\infty)} .
\label{angdefBcc}
\end{equation}
while for the Melvin branch we get $M$ and $W$ in terms of the central magnetic
field:
\begin{eqnarray}
G M= \frac{\frac{\gamma}{\alpha} B(0)-1}{3(1-{\gamma\over 6})}   \\
G W=-\frac{\frac{\gamma}{2\alpha} B(0)+1-{\gamma\over 4}}{3(1-{\gamma\over 6})}.
\label{MWdeponBcc}
\end{eqnarray}
Note that although the relation between the central magnetic field and the
angular deficit is modified, the critical magnetic field is still given by the
minimal coupling expression, (\ref{critB}).

We thus expect to find in the case of conformal coupling, cosmic string
solutions that do not differ significantly from the minimally coupled ones.
This will indeed be the main conclusion but there is one important exception:
the  Bogomolnyi cosmic strings. As mentioned above, this is the case where
there is a complete balance between the repulsive and attractive contributions
(to ${\cal T}_{\mu \nu}$) which is reflected by the fact that $N(x)=1$ is a
consistent solution. The straightforward way to study the analogous situation
for conformal coupling  is to examine whether the system still admits $N(x)=1$
as a solution. It turns out that this is not the case, and $N(x)=1$ is
inconsistent with the field equations. One may arrive at this  conclusion by a
direct computation of the derivatives of $N(x)$ at $x=0$ up to order $p$ 
(exploiting all the field equations) for any $\alpha$ and $\gamma$ and studying
the possibility that they all vanish simultaneously. It is found that $p$
cannot be greater than $9$, namely, there cannot exist a solution where all the
first $10$ derivatives of $N(x)$  at $x=0$ vanish. This evidently excludes the
possibility of a constant $N(x)$. Since high order derivatives are involved,
the algebraic  manipulations are rather cumbersome and they will not be shown
here explicitly.

Further study of the nature of the solutions is possible only by numerical
analysis. The results are described in the next section.

\section{Numerical Results and Conclusion} \setcounter{equation}{0}

In this section the essential characteristics of the cosmic strings with
conformal  coupling to gravity are presented, mainly in a graphical manner. For
comparison, the corresponding characteristics for minimal coupling are also
shown. The results were obtained by the MATHEMATICA package using an iterative
procedure for solving the four coupled field equations. In order to insure the
consistency  of the results several checks were made using the constraints and
relations like (\ref{constraintccdim}) and (\ref{WBcc}) and their minimal
coupling analogues.

\begin{figure}[ht]
\epsfxsize=0.8\textwidth
\epsfbox{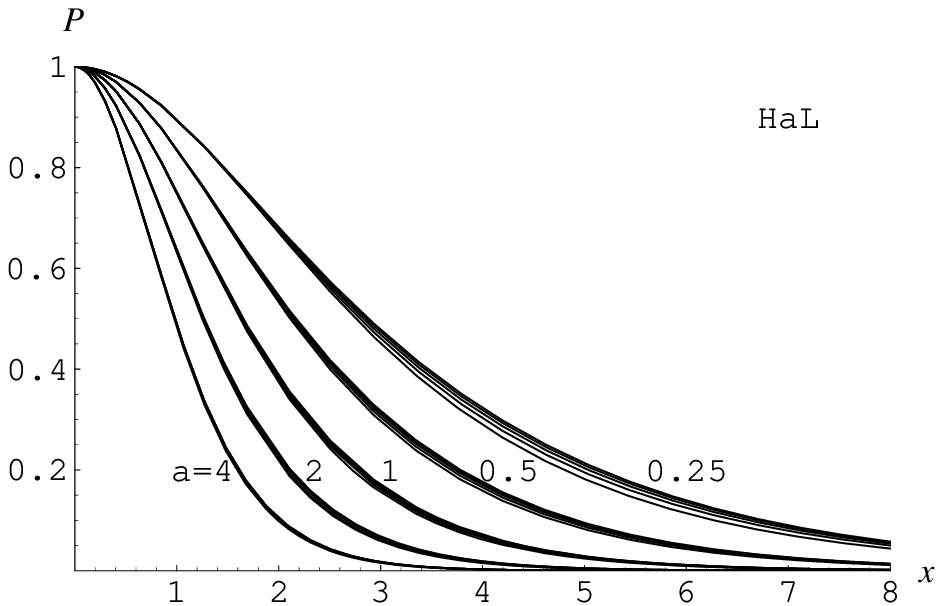}
\vspace{2cm}
\epsfxsize=0.8\textwidth
\epsfbox{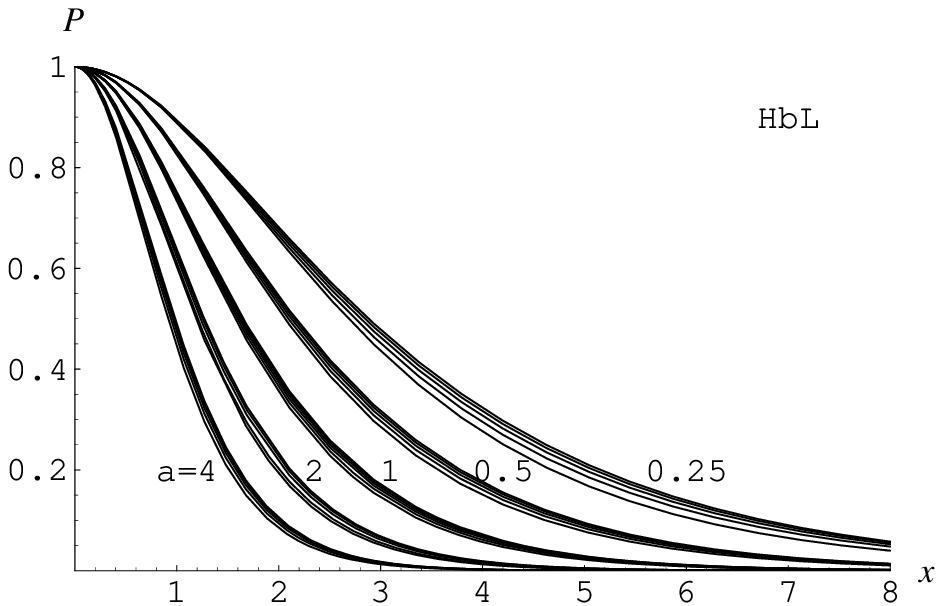}
\caption{\small The magnitude of the gauge field $P(x)$ for
$\alpha$=0.25, 0.5, 1, 2, 4 and several values of $\gamma$ - see text.
\protect\newline
(a) minimal coupling. (b) conformal coupling. }
\label{figp}
\end{figure}

The scalar and gauge field look generally as in flat space. The gauge field, or
$P(x)$ is sensitive to $\alpha$ but does not depend very much on $\gamma$. It
slightly concentrates toward the axis when  $\gamma$ is increased.
Fig.\ref{figp} depicts $P(x)$ for $\alpha=0.25, 0.5, 1, 2, 4$. For each value
of $\alpha$, $\gamma$ takes values in steps of $0.25$ starting from $\gamma=0$
(flat space flux tube) up to a maximal value which depends on $\alpha$: $1, 1,
1.25, 1.75, 2$ for the above values respectively. Altogether there are 33
points in the $\alpha$-$\gamma$ plane represented in the figure.  
The fact that the vector field concentrates while $\alpha$ increases, results
from the fact that $\alpha$ represents the ratio between the scalar and the
vector characteristic length scales.   
One should however keep in mind that the behavior with respect to "real"
distance, $r=x/\sqrt{\lambda v^2}$ may be different since changes in $\alpha$
and $\gamma$ are associated with changes in $\lambda$ and $v$ and an increase
of those tends to expand $P(r)$. 

\begin{figure}[ht]
\epsfxsize=0.8\textwidth
\epsfbox{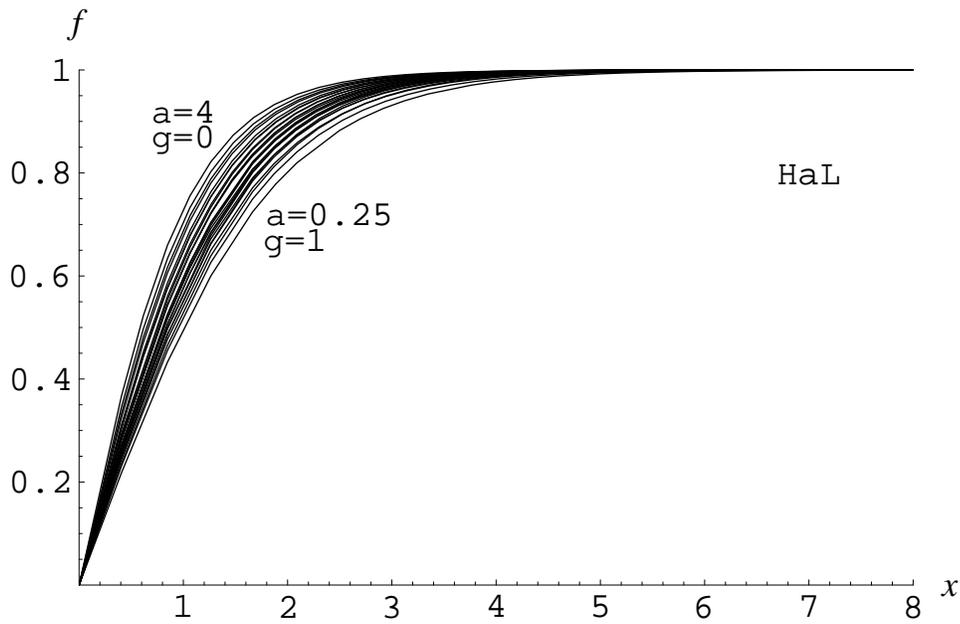}
\vspace{2cm}
\epsfxsize=0.8\textwidth
\epsfbox{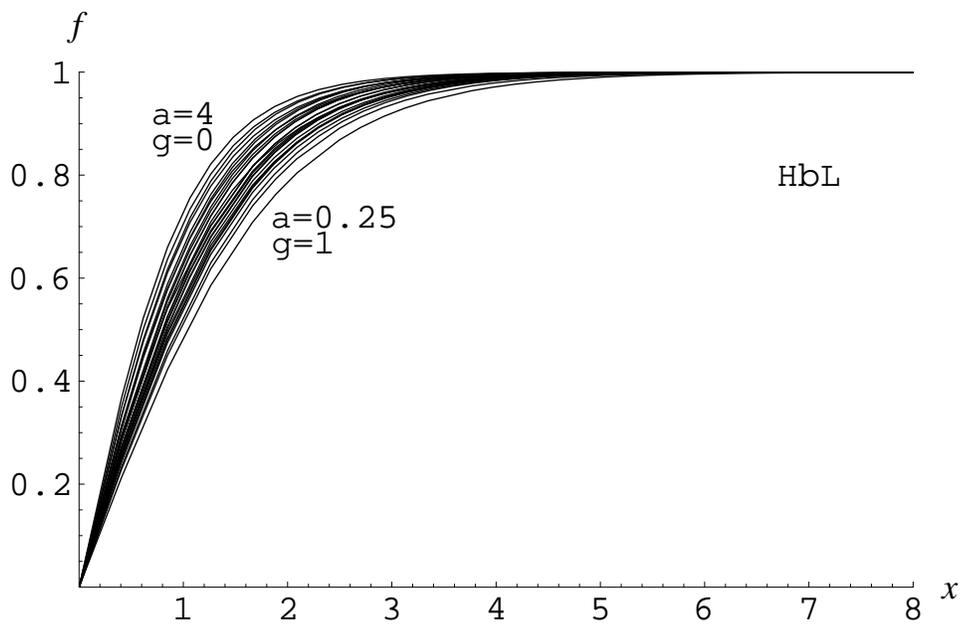}
\caption{\small The magnitude of the scalar field $f(x)$ for
$\alpha$=0.25, 0.5, 1, 2, 4 and several values of $\gamma$ - see text. 
The "envelope" corresponds to the values shown. 
(a) minimal coupling. (b) conformal coupling. }
\label{figf}
\end{figure}

The scalar field is quite insensitive to changes in both $\alpha$ and $\gamma$
as can be learnt from Fig.\ref{figf}: An increase of $\alpha$ and decrease of
$\gamma$ have a similar weak effect of increasing the rate of approach of $f(x)$
towards $1$. Equivalently, the slope of the scalar field on the axis increases
with $\alpha$ and decreases with $\gamma$. The $f(x)$ curves correspond to the
same 33 points in the  $\alpha$-$\gamma$ plane mentioned above.

\begin{figure}[ht]
\epsfxsize=0.8\textwidth
\epsfbox{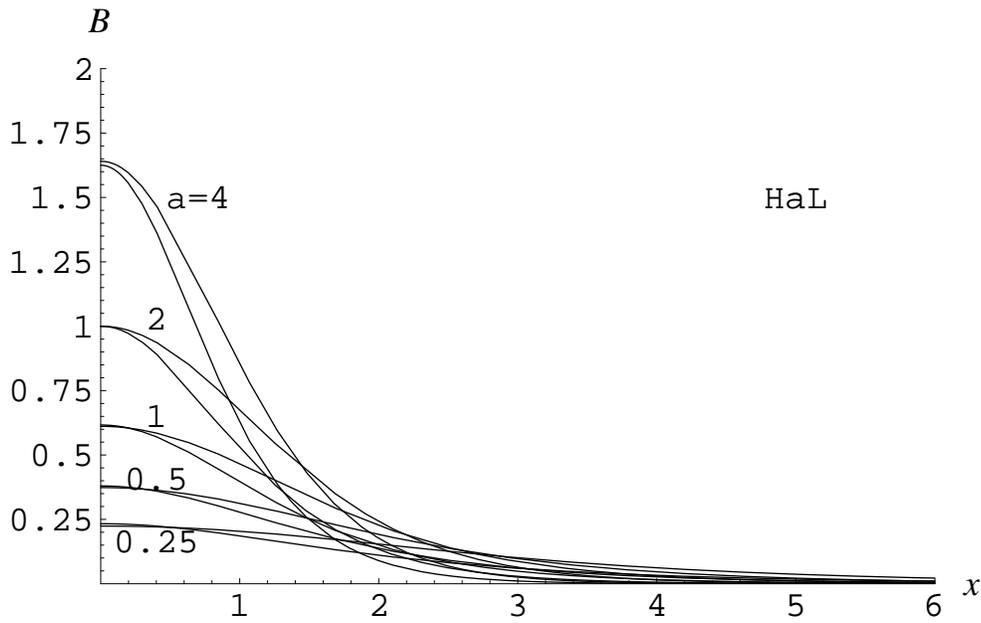}
\vspace{2cm}
\epsfxsize=0.8\textwidth
\epsfbox{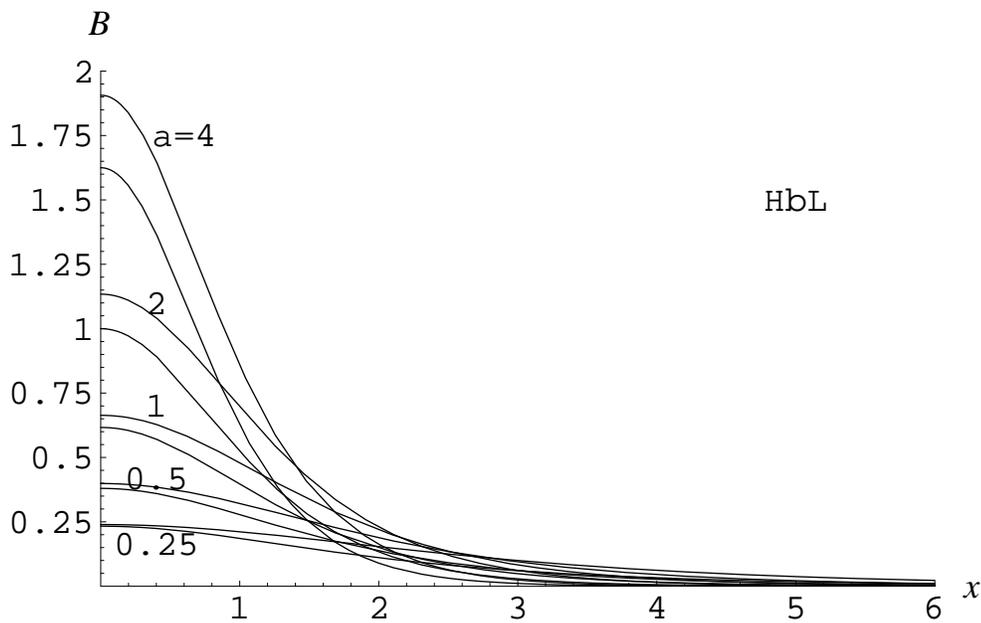}
\caption{\small The magnitude of the magnetic field $B(x)$ for
$\alpha$=0.25, 0.5, 1, 2, 4. Two curves are shown for each value of $\alpha$: 
$\gamma$=0 (the lower) and respectively $\gamma$= 1, 1, 1.25, 1.75, 2. 
(a) minimal coupling. 
(b) conformal coupling.}
\label{figb}
\end{figure}

Fig.\ref{figb} presents the behavior of the magnetic field. For the sake of
clarity only the extreme values of $\alpha$ and $\gamma$ appear. The general
tendency is that the magnetic field increases with both $\alpha$ and $\gamma$.
However, this tendency reverses for large $x$ since $B(x)$ (as well as the
other fields) approaches faster its vacuum value for larger values of $\alpha$.
There is also a slight decrease of the magnetic field on the axis in the
minimally coupled string for $\alpha <2$.     

\begin{figure}[ht]
\epsfxsize=1.0\textwidth
\epsfbox{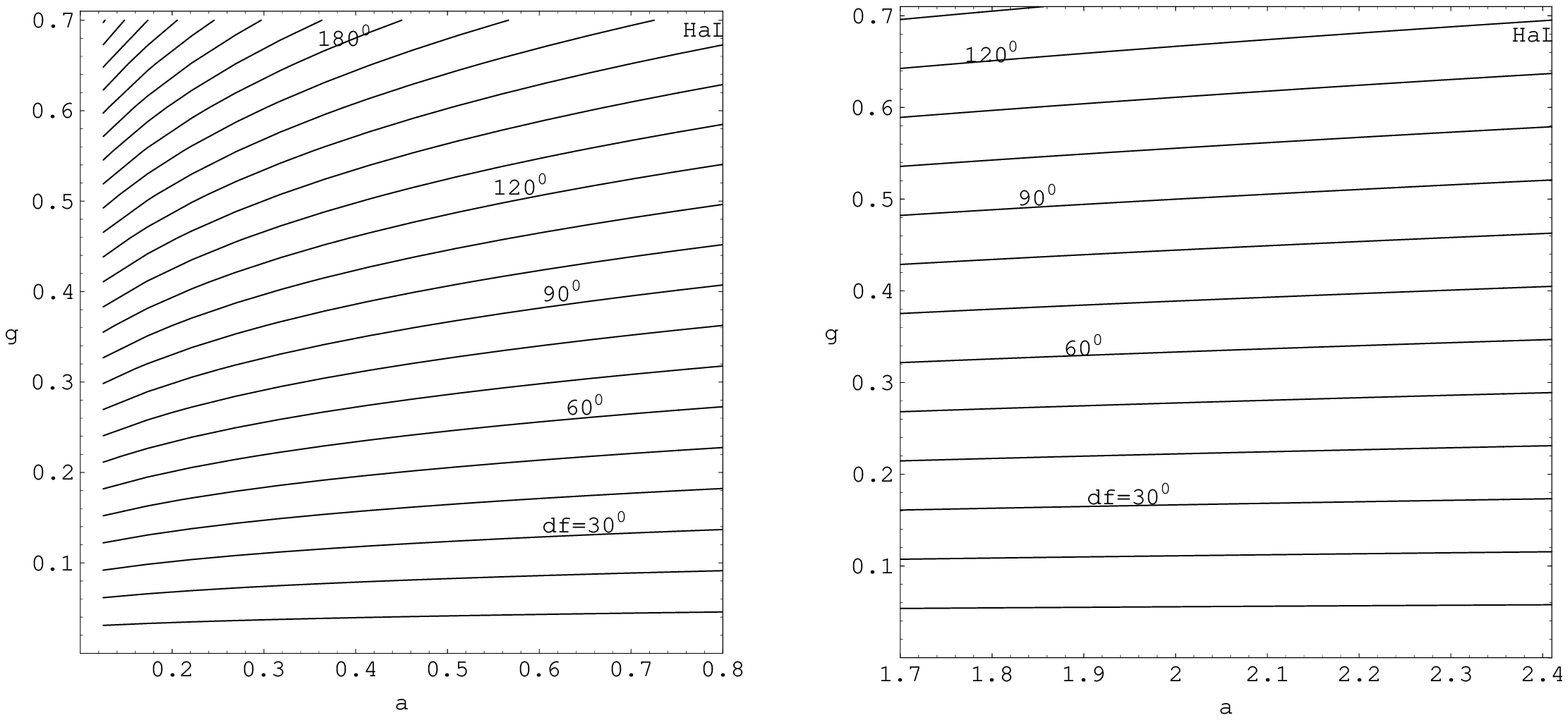}
\vspace{2cm}
\epsfxsize=1.0\textwidth
\epsfbox{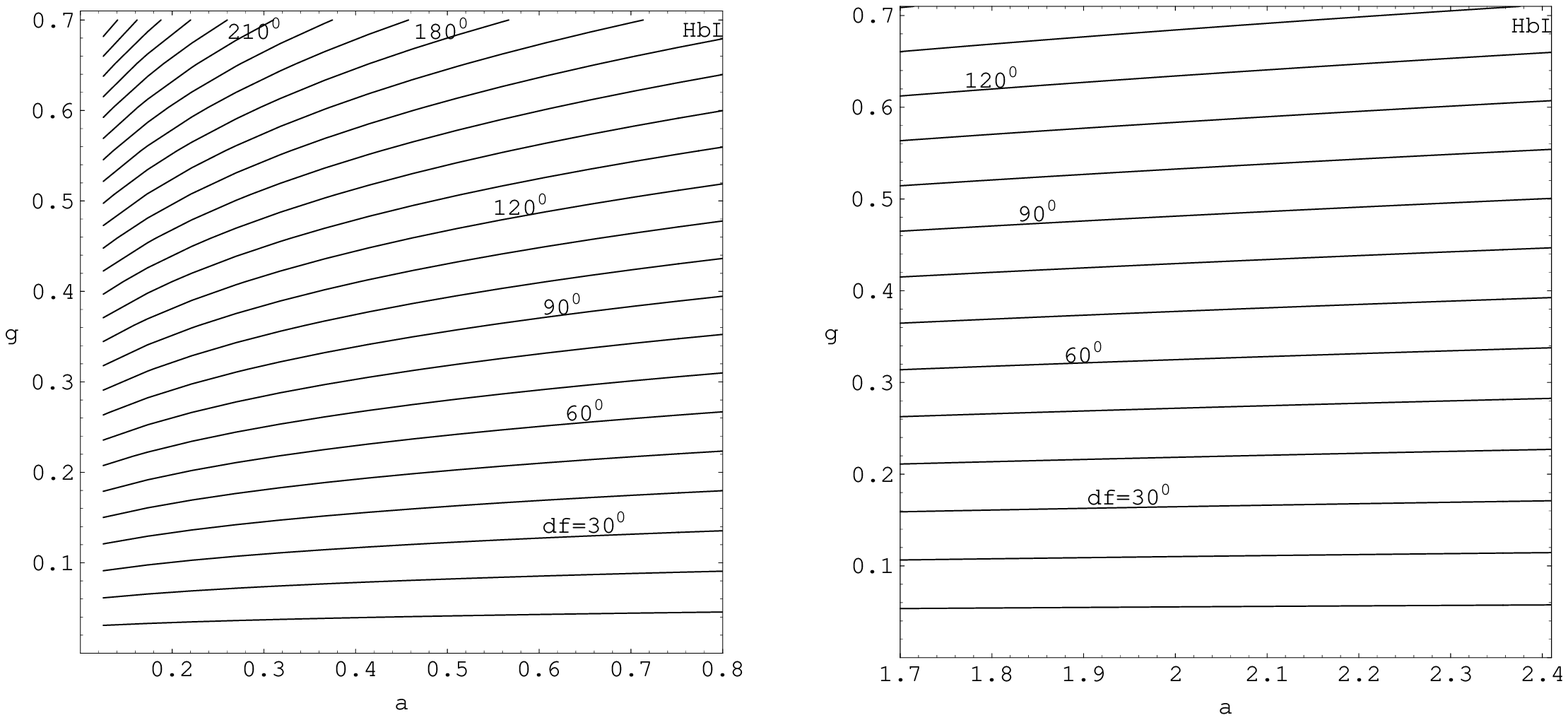}
\caption{\small Curves of constant angular deficit $\delta\phi$ in two 
portions of the $\alpha$-$\gamma$ plane. The curves are spaced in steps of 
$10^0$. (a) minimal coupling. (b) conformal coupling.}
\label{figdeltaphi}
\end{figure}

The metric components are more sensitive to changes in $\alpha$ and $\gamma$
and there is a boundary in the $\alpha$-$\gamma$ plane beyond which no open
solutions exist at all.  A complete mapping of the $\alpha$-$\gamma$ plane is a
matter for further analysis. A general idea as of the  shape of this boundary
(i.e. along which $\delta\phi=2\pi$) may be obtained however from Fig.
\ref{figdeltaphi}.  

\begin{figure}[ht]
\epsfxsize=0.8\textwidth
\epsfbox{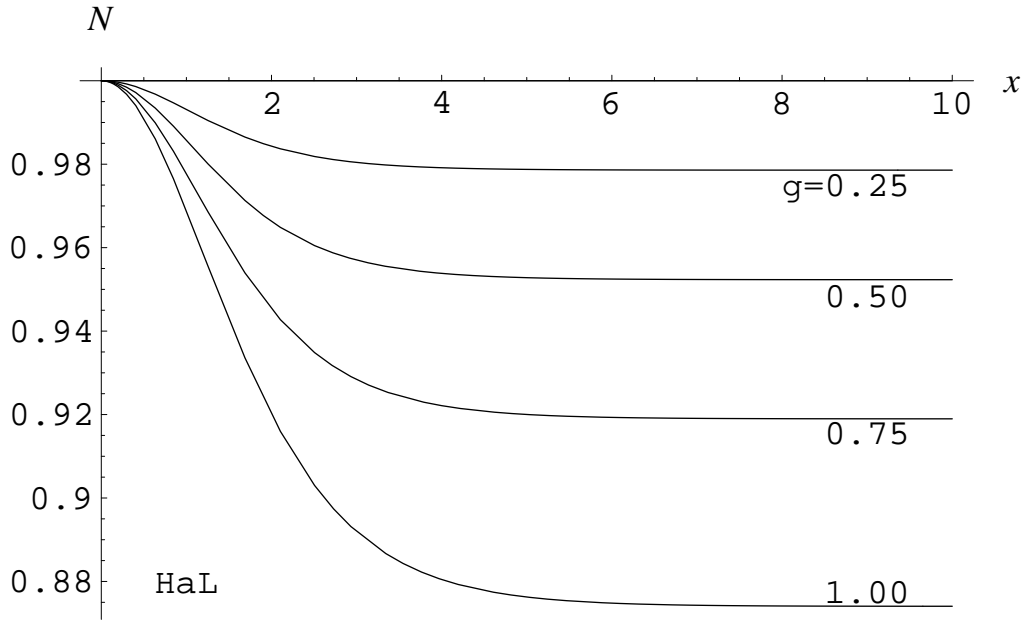}
\vspace{2cm}
\epsfxsize=0.8\textwidth
\epsfbox{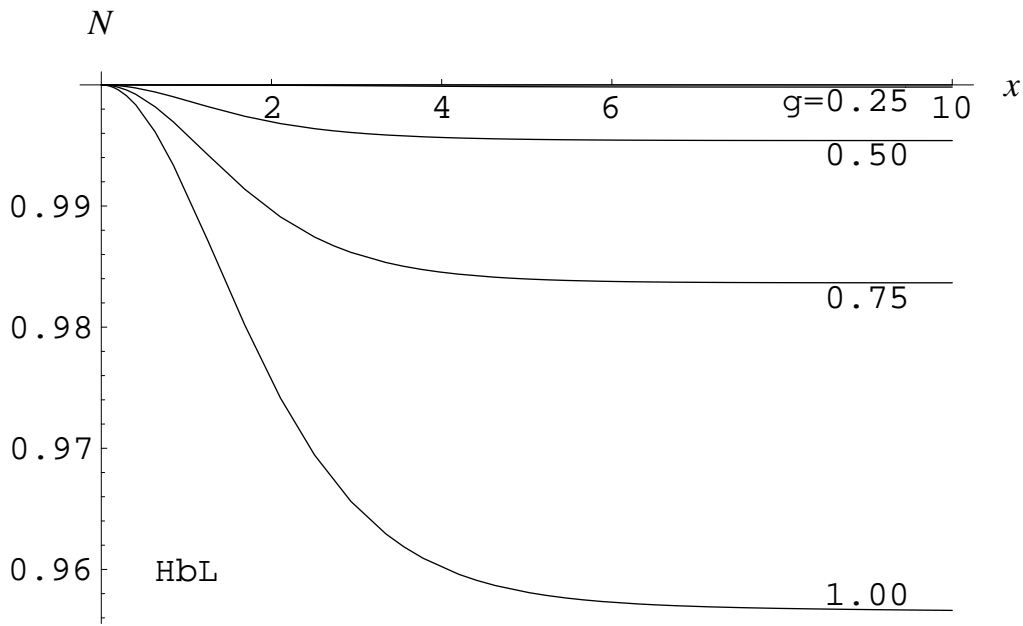}
\caption{\small The magnitude of the metric component $N(x)$ for
$\alpha$=0.25 and $\gamma$=0, 0.25, 0.50, 0.75, 1.00. \protect\newline 
(a) minimal coupling. 
(b) conformal coupling. Note the difference in scale. 
$\gamma$=0 corresponds to $N(x)=1$; for $\gamma$=0.25 the deviation with respect
to $N(x)=1$ is beyond the resolution of the graph.}
\label{fign0}
\end{figure}

\begin{figure}[ht]
\epsfxsize=0.8\textwidth
\epsfbox{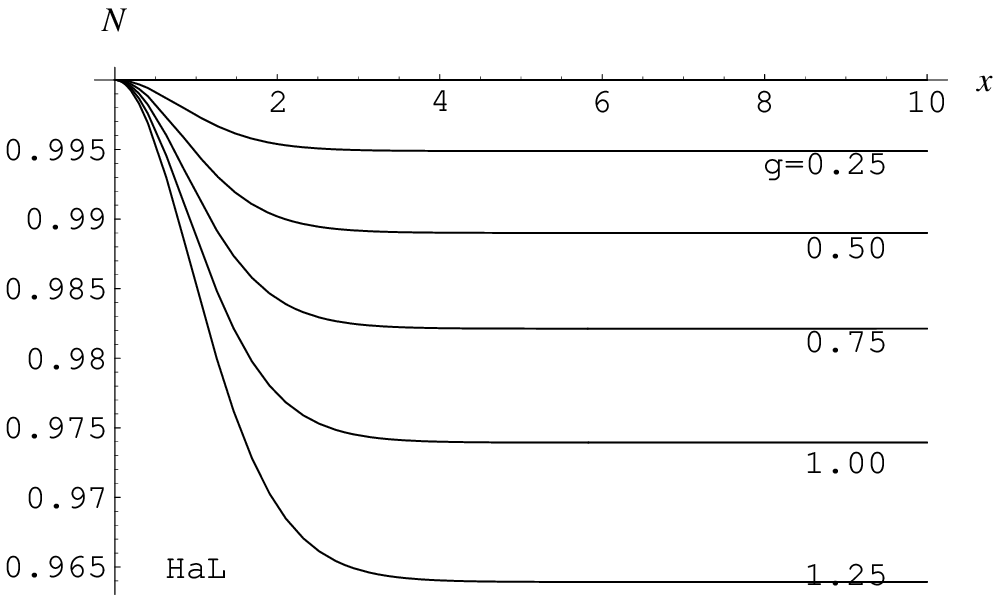}
\vspace{2cm}
\epsfxsize=0.8\textwidth
\epsfbox{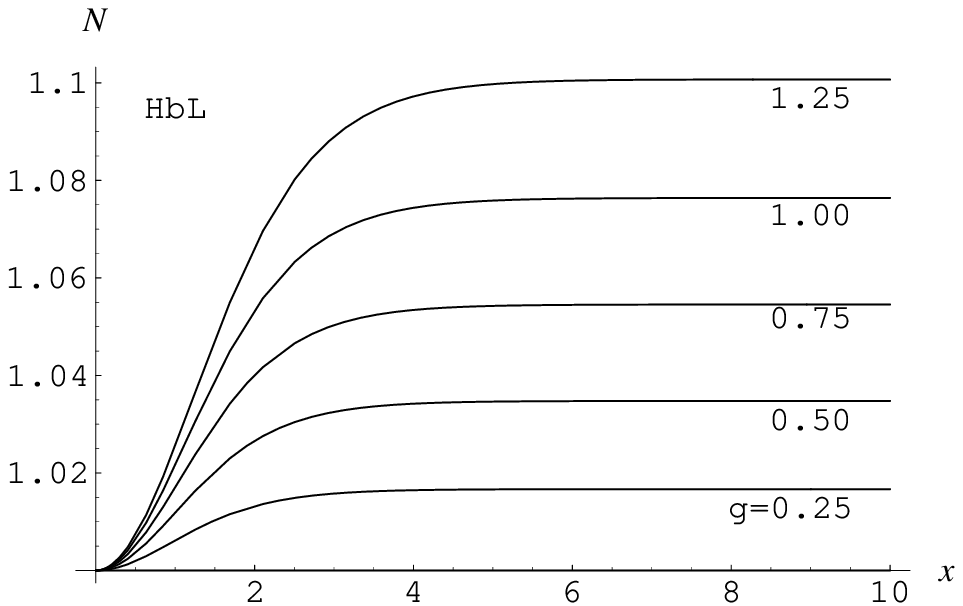}
\caption{\small The magnitude of the metric component $N(x)$ for
$\alpha$=1 and $\gamma$=0, 0.25, 0.50, 0.75, 1.00, 1.25. \protect\newline 
(a) minimal coupling. 
(b) conformal coupling. Note the difference in scale. $\gamma$=0 corresponds 
to $N(x)=1$. }
\label{fign2}
\end{figure}

\begin{figure}[ht]
\epsfxsize=0.8\textwidth
\epsfbox{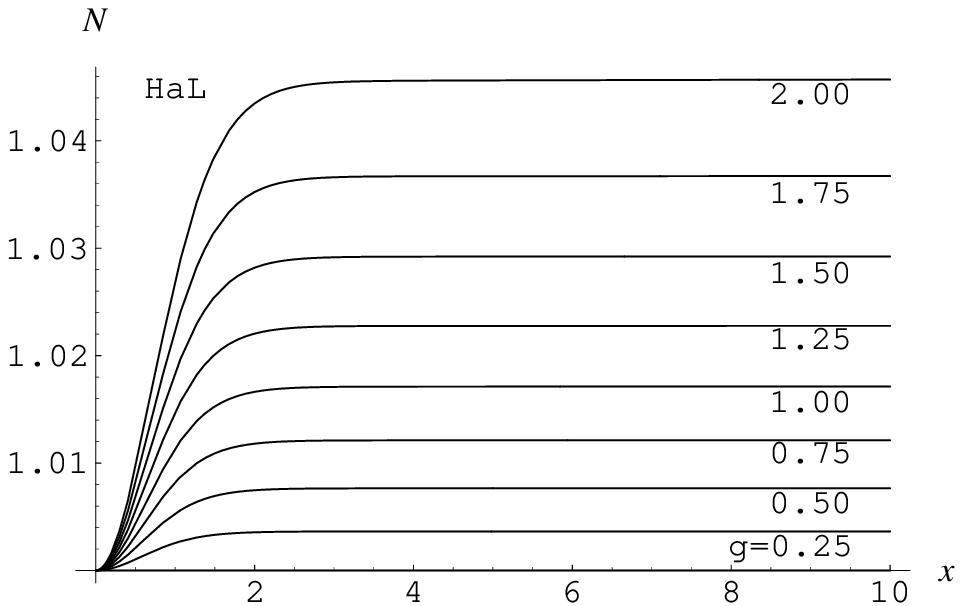}
\vspace{2cm}
\epsfxsize=0.8\textwidth
\epsfbox{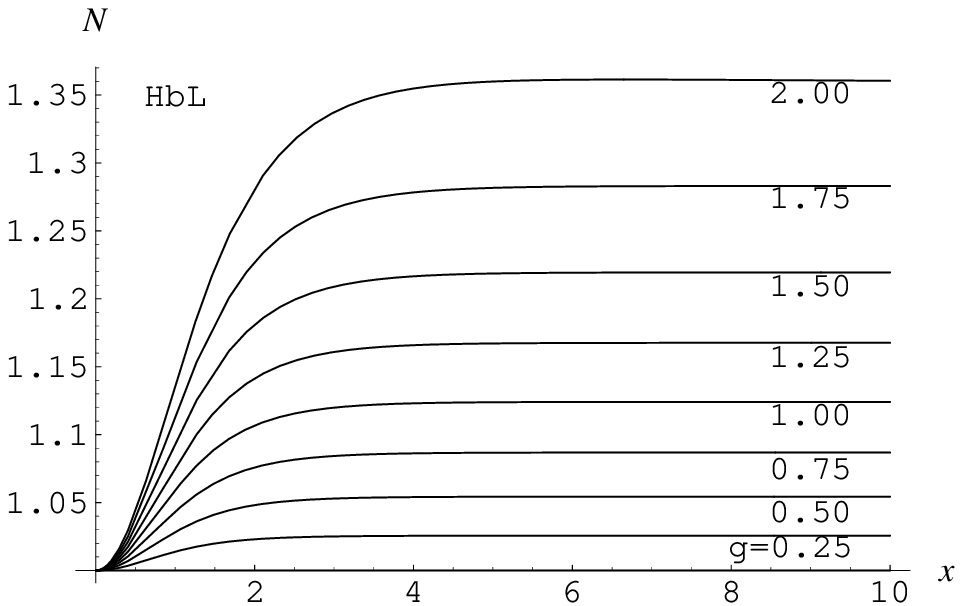}
\caption{\small The magnitude of the metric component $N(x)$ for
$\alpha$=4 and $\gamma$=0, 0.25, 0.50, 0.75, 1.00, 1.25, 1.50, 1.75, 2.00. 
(a) minimal coupling. 
(b) conformal coupling. Note the difference in scale. 
$\gamma$=0 corresponds to $N(x)=1$. }
\label{fign4}
\end{figure}

The conformal coupling indirectly intensifies the gravitational effects with
respect to the case of minimal coupling. The temporal metric component $N(x)$
changes its behavior significantly; especially for $\alpha<2$ where $N(x)$ is a
decreasing function for all $\gamma$ in the minimally coupled case but not so
for conformal coupling. For  $\alpha \geq 2$, where the general form of $N(x)$
is similar for both cases, the red/blue-shift factor $N(\infty)$ is much larger
for conformal coupling than for minimal coupling. These features are evident
from Fig.\ref{fign0}-Fig.\ref{fign4} which present $N(x)$ for several
representative points in the $\alpha$-$\gamma$ plane.

\begin{figure}[ht]
\epsfxsize=0.8\textwidth
\epsfbox{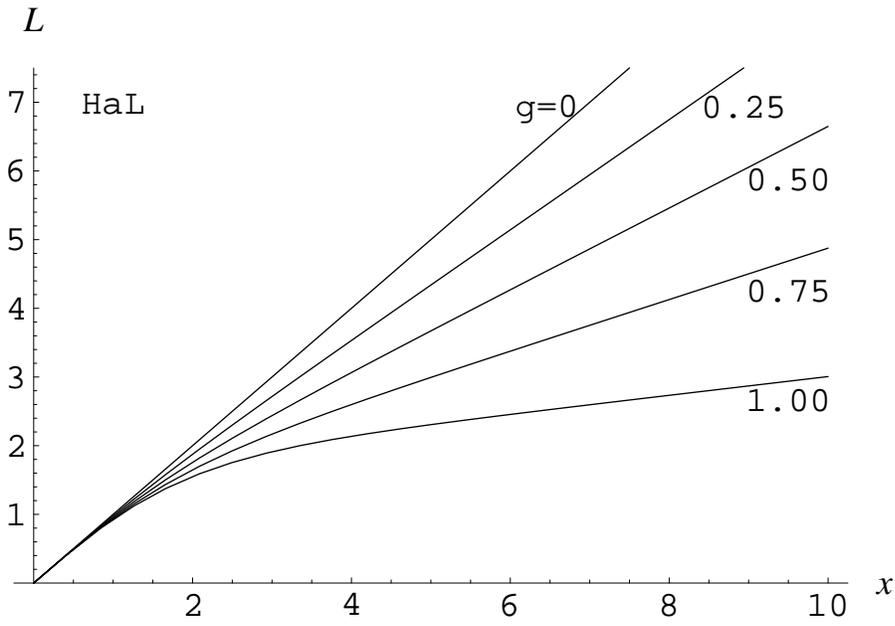}
\vspace{2cm}
\epsfxsize=0.8\textwidth
\epsfbox{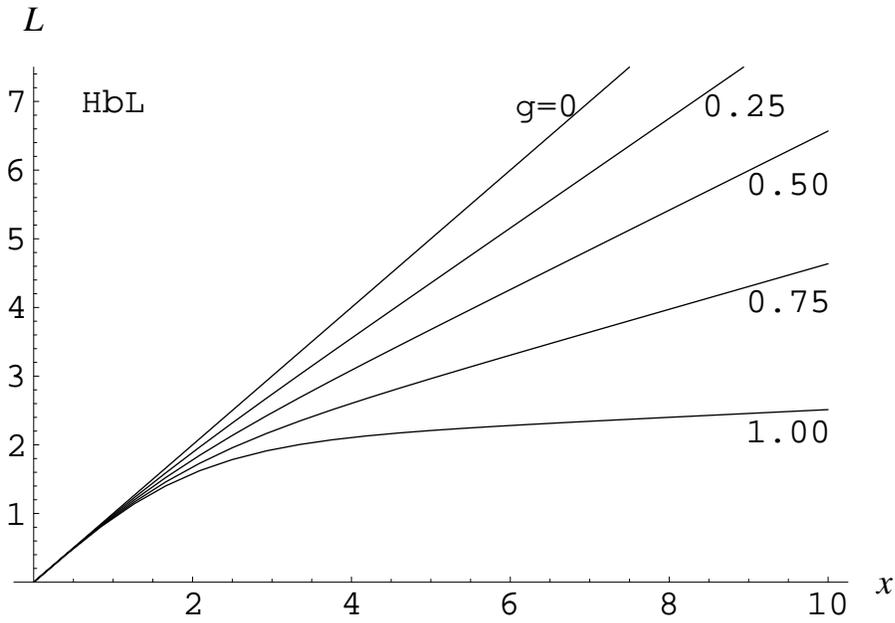}
\caption{\small The magnitude of the metric component $L(x)$ for
$\alpha$=0.25 and $\gamma$=0, 0.25, 0.50, 0.75, 1.00. \protect\newline 
(a) minimal coupling. 
(b) conformal coupling.}
\label{figl0}
\end{figure}

\begin{figure}[ht]
\epsfxsize=0.8\textwidth
\epsfbox{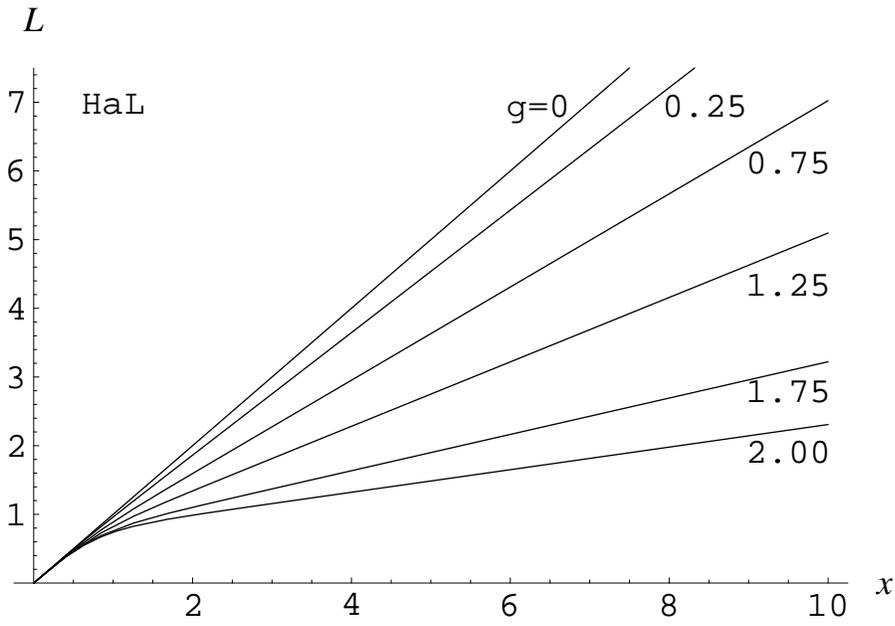}
\vspace{2cm}
\epsfxsize=0.8\textwidth
\epsfbox{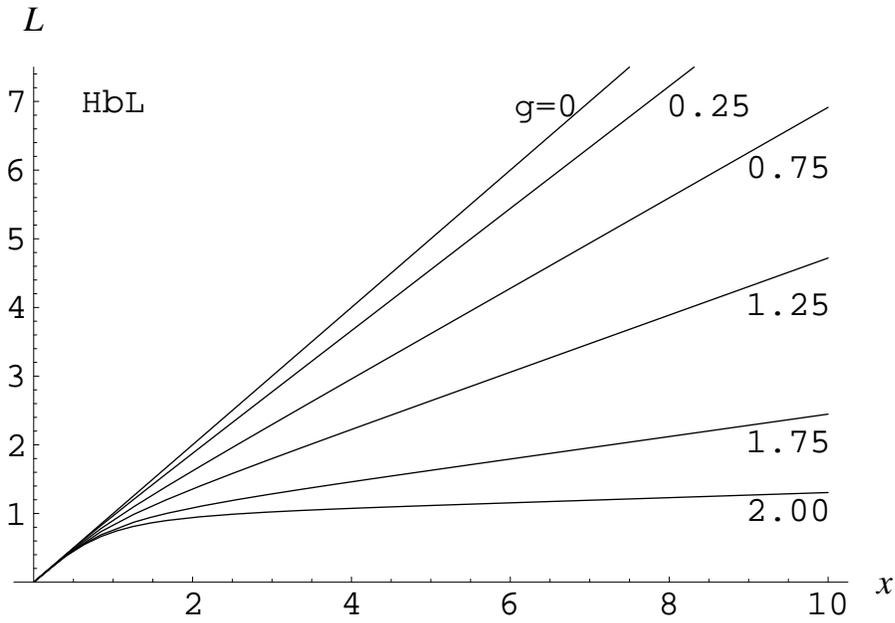}
\caption{\small The magnitude of the metric component $L(x)$ for
$\alpha$=4 and $\gamma$=0, 0.25, 0.75, 1.25, 1.75, 2.00. \protect\newline 
(a) minimal coupling. 
(b) conformal coupling.}
\label{figl4}
\end{figure}

The most prominent feature of the cosmic string solutions is the angular
deficit $\delta\phi$ which is in one-to-one correspondence with the asymptotic
value of the slope of the angular metric component $L(x)$.
Fig.\ref{figl0}-Fig.\ref{figl4} show this function and short inspection shows
that $\delta\phi$ decreases with $\alpha$ and increases with $\gamma$ in both
types of coupling. The angular deficit is always larger in the conformally
coupled case.

\begin{figure}[ht]
\epsfxsize=1.0\textwidth
\epsfbox{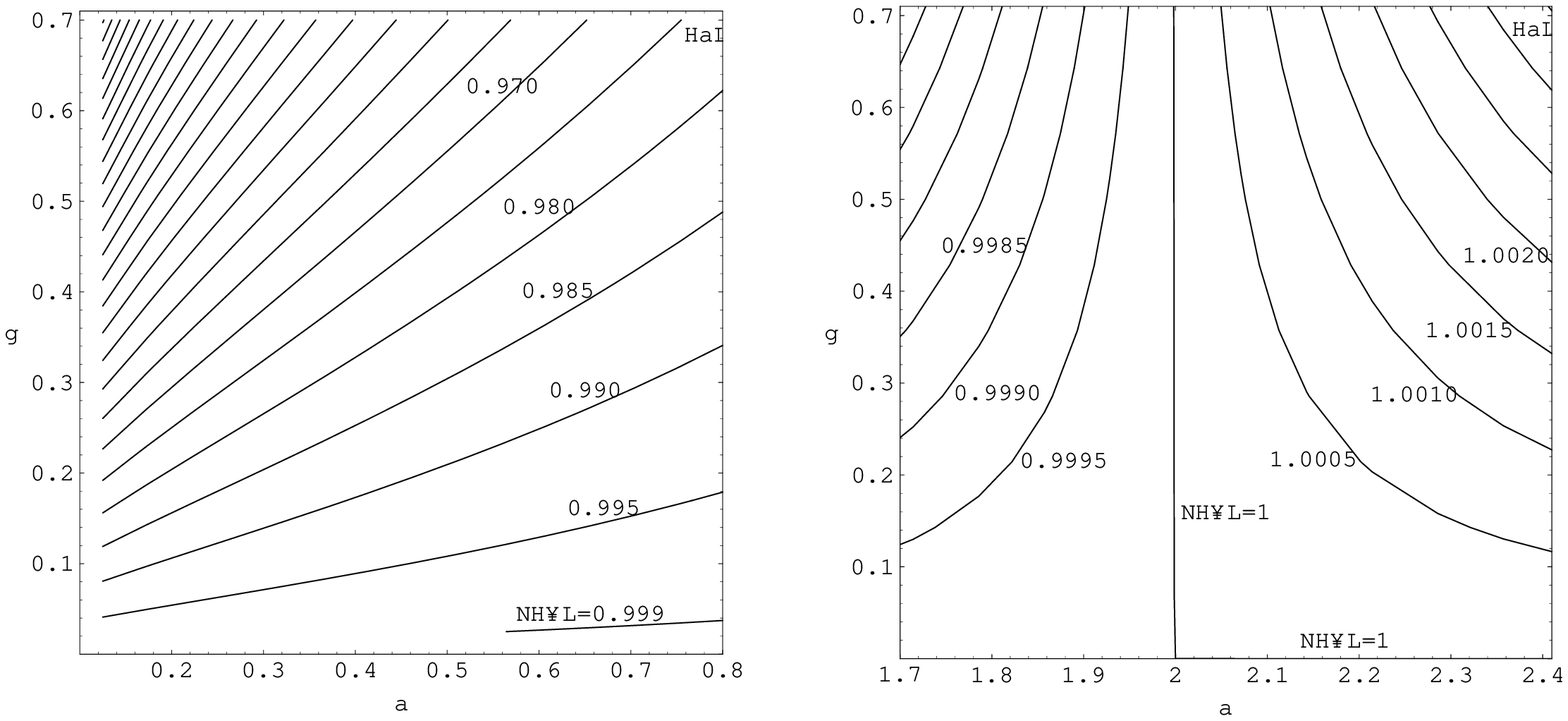}
\vspace{2cm}
\epsfxsize=1.0\textwidth
\epsfbox{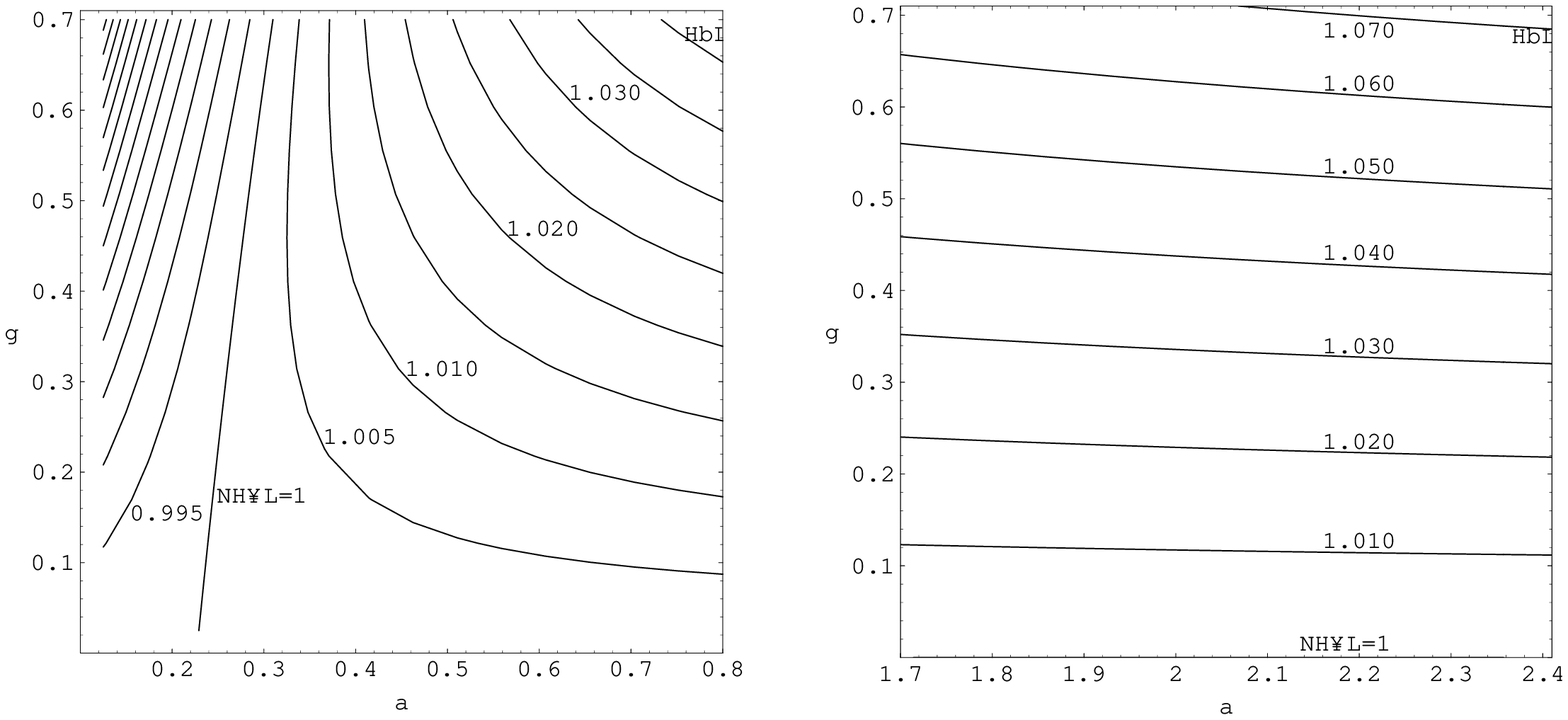}
\caption{\small Curves of constant red/blue-shift factor $N(\infty)$ in two 
portions of the $\alpha$-$\gamma$ plane. 
Note the non-equal spacing of the curves. 
(a) minimal coupling. (b) conformal coupling.}
\label{figni}
\end{figure}

The $\alpha$ and $\gamma$ dependence of these two characteristics $N(\infty)$ 
and $\delta\phi$ is more clearly seen in Fig.\ref{figni} and
Fig.\ref{figdeltaphi} which show a portion of the $\alpha$-$\gamma$ plane. Note
the very different behavior of the red/blue-shift factor $N(\infty)$ in both
cases. For minimal coupling the line $\alpha =2$ is the Bogomolnyi line which
separates between $N(\infty)<1$ and $N(\infty)>1$. A similar $N(\infty)=1$ line
exists for the conformal coupling; however as the non-existence of Bogomolnyi
strings implies, it does not correspond to solutions with $N(x)=1$ as happens
for minimal coupling, but merely to the case of no red/blue-shift between
infinity and the core of the string.

the abelian Higgs cosmic strings with conformal coupling seem therefore to be as
respectable as their minimally coupled counterparts. Both types share the same
general tendencies and typical characteristics and most of the investigation
may be done for both cases along parallel paths. The only significant
difference is the non-existence of the Bogomolnyi cosmic strings in the case of
conformal coupling. 

\vspace{1cm}
\noindent
{\Large \bf Acknowledgements}
\vspace{0.5cm}

I am grateful to R. Brustein, A. Davidson, A.L. Larsen, R. Madden and N.K.
Nielsen for useful discussions and correspondence.


\newpage

\newpage

\newpage

\newpage

\newpage

\newpage

\newpage

\newpage

\newpage

\newpage

\end{document}